\documentclass[preprint, 3p]{elsarticle}

\usepackage{ecrc}

\usepackage[hidelinks]{hyperref}

\usepackage{comment} 

\volume{}

\firstpage{1}

\runauth{Flohr, Schuß, Wallach, Krüger, \& Riener}

\usepackage{amssymb}

\usepackage[figuresright]{rotating}

\usepackage[T1]{fontenc}

\begin{document}

\begin{frontmatter}




\title{Designing for Passengers’ Information Needs on Fellow Travelers: \\A Comparison of Day and Night Rides \\ in Shared Automated Vehicles}


\author[ES,UdS,1]{Lukas A. Flohr}
\address[ES]{Ergosign GmbH, Saarbr{\"u}cken \& Munich, Germany}
\address[UdS]{Saarbr{\"u}cken Graduate School of Computer Science, Saarland Informatics Campus, Saarbr{\"u}cken, Germany}
\footnote[1]{Both authors contributed equally to the paper.}

\author[THI,JK,1]{Martina Schu{\ss}}
\address[THI]{Technische Hochschule Ingolstadt (THI), Ingolstadt, Germany}
\address[JK]{Johannes Kepler Universit\"at, Linz, Austria}

\author[HSKL,ES]{\\Dieter P. Wallach}
\address[HSKL]{University of Applied Sciences Kaiserslautern, Kaiserslautern, Germany}

\author[DFKI]{Antonio Kr{\"u}ger}
\address[DFKI]{German Research Center for Artificial Intelligence (DFKI), Saarland Informatics Campus, Saarbr{\"u}cken, Germany}

\author[THI]{Andreas Riener}

\begin{abstract}
Shared automated mobility-on-demand promises efficient, sustainable, and flexible transportation. Nevertheless, security concerns, resilience, and their mutual influence – especially at night – will likely be the most critical barriers to public adoption since passengers have to share rides with strangers without a human driver on board. 
As related work points out that information about fellow travelers might mitigate passengers' concerns, we designed two user interface variants to investigate the role of this information in an exploratory within-subjects user study ($N=24$). Participants experienced four automated day and night rides with varying personal information about co-passengers in a simulated environment. The results of the mixed-method study indicate that having information about other passengers (e.g., photo, gender, and name) positively affects user experience at night. In contrast, it is less necessary during the day. Considering participants' simultaneously raised privacy concerns, balancing security and privacy demands poses a substantial challenge for resilient system design.
\end{abstract}


\end{frontmatter}



\section{Introduction}
The rapid progress of automated driving technologies promises to revolutionize public transportation (PT) by creating automated mobility-on-demand (AMoD, \cite{Pavone2016}) systems. In AMoD, passengers are transported by driverless vehicles – i.e., cars with SAE level 4 (high driving automation) or level 5 (full driving automation) capabilities \cite{SAEInternational2021}. Those automated vehicles (AVs) will be guided by intelligent traffic management systems. They enable efficient route planning and smart ride-sharing, which will decrease the number of vehicles on the streets \cite{Spieser2014}, turning traffic jams into a remembrance of the past \cite{cox2016autonomous}. While it is unclear whether private vehicles or shared options will dominate in the future, the (potentially) most beneficial scenarios unfold when shared mobility prevails over private vehicle ownership \cite{litman2017autonomous,papa2018sustainable}.
To arrive at its full potential, it is imperative that passengers not only share vehicles but also share rides \cite{currie2018lies}. Automated ride-sharing has many similarities with today's forms of PT. However, its demand-orientated service approach offers both high(er) efficiency and comfort for passengers. Shared AMoD (SAMoD) will lead to fewer cars in the streets \cite{itf2015urban_mobility} and -- as simulations suggest \cite{Spieser2014} -- lower air pollution \cite{thomopoulos2015autonomous}. 
Since drivers are not needed anymore, 
passengers will solely interact with the AVs and the intelligent traffic management systems. Two considerations have to be addressed when developing such systems. First, passengers need to rely on digital user interfaces (UIs) that provide them with trustful information in real-time, as well as with efficient controls for their travel needs. Second, in SAMoD systems, passengers will share rides with strangers during both day and night times. Since there is no human authority (e.g., a bus driver) present anymore in AVs, users' acceptance is likely to be influenced by the presence of co-passengers \cite{salonen2019towards, schuss2021letssharearide}. Information about fellow travelers prior to and during the ride seems to have the potential to affect user acceptance positively \cite{koenig_generationY}. 
The question arises whether these findings of are further influenced by gender and time of the day as, e.g., \cite{schuss2021letssharearide} found that especially women are feeling anxious to share rides at night times. Understanding the potential influence of gender and time of the day on user acceptance, perceived security, and resilience in SAMoD systems is crucial for designing inclusive and user-centered transportation solutions. 

    To overcome acceptance challenges and design resilient SAMoD systems, more suitable service concepts and UIs are required. We agree with \cite{schuss-feministHCI-journal} that "research is needed on how to enhance women's security while not leaving other groups of people out". 
    Research in the context of PT shows that feelings of anxiety and unease when traveling with strangers greatly influence perceived security \cite{currie2013factors}. 
    Building upon related work, we hypothesize that knowledge about co-passengers might positively affect perceived security, acceptance, trust, and overall user experience (UX).
    Consequently, this work investigates the following research question (RQ):
    \begin{itemize}
        \item[] \textbf{RQ.} How are SAMoD passengers' perceptions of security and corresponding UX, trust, acceptance, and emotions influenced by the time of day of a shared ride and their knowledge of co-passengers?
    \end{itemize}
    Most related work mentioned above used (online) surveys to examine user acceptance of (shared) AVs. As such, participants lack experience using SAMoD systems. Therefore, we investigate our research question in a user study conducted with 24 participants in Germany using a controlled simulated environment enabling this very experience. On this basis, our empirically grounded findings confirm but also challenge previous works on shared AVs and shed new light on the effects of time of day and the interrelation of (co-)passengers. The paper informs on SAMoD passengers' information needs, such as information about fellow travelers but also points out that providing such information may come with tensions and costs, e.g., reduced privacy. The findings 
     may help SAMoD service providers create resilient systems and desirable interactions in shared AVs that facilitate the technology's adaption and help passengers feel secure.
\section{Background and Related Work}
From an HCI perspective, SAMoD faces significant acceptance barriers. In this section, we provide an overview of factors influencing the acceptance of SAMoD and discuss methods for designing and evaluating adequate UIs to tackle those barriers. Finally, we derive our research question and situate the contributions of this paper among related work.

\subsection{Factors Influencing the Acceptance of Shared Automated Mobility} 
\label{related-wor-acceptance}
The significance of shared mobility modes will continue to grow \cite{currie2018lies, sperling2018three-revolutions} and public entities must identify opportunities to engage with these to ensure their benefits are widely and equitably shared. However, public acceptance poses major challenges for AVs \cite{Kaur2018}. Acceptance factors identified by related work \cite{Kaur2018, Lundquist1218637, salonen2019towards, frison2017driving, wintersberger2018man} can generally be allocated to overall demographic factors, system reliability and performance issues, security concerns, users' expectations and trust, and privacy concerns vs. information demands.

Recent studies examined people's willingness to switch from private to shared mobility modes in the future of mobility \cite{schuss2021letssharearide}, with demographics such as gender \cite{SALONEN2018106} and age \cite{KRUEGER2016343} as predictors for the adoption of shared AVs (SAVs) and young men being the group with the highest openness towards this technology. 
Some of the most critical barriers to accepting SAVs are related to security concerns because rides will have to be shared with strangers \cite{sarriera_to-share-or-not, schuss-feministHCI-journal, schuss2021letssharearide, piao2016public, sanguinetti2019ok, salonen2019towards}. 
Clayton et al. \cite{2020_Clayton_whowillusethemandwilltheyshare} examined the willingness to share an AV with other passengers and found uncertainty about sharing and a strong preference for privately owned vehicles. An online survey by Pakusch et al. \cite{pakusch2018unintended} underlines the reluctance to switch from private rides to shared ones, predicting that private AVs will dominate the future of automated driving. Mapping these results to the context of SAVs, Lavieri and Bhat \cite{LAVIERI2019242} found that people were willing to pay extra fees for trips in SAVs when only the vehicle, but not the trip, is shared with others. Their study indicated that privacy and security concerns would prevent participants from opting for sharing rides with strangers — affecting commuting trips to a minor extent than rides for pleasure purposes \cite{LAVIERI2019242}.

A qualitative user study by Schuß et al. \cite{schuss2021letssharearide, schuss-feministHCI-journal} emphasizes security concerns as an important issue for automated ride sharing, especially for women — and particularly during the night. These results were confirmed by Piao et al. \cite{piao2016public, SALONEN2018106}. Passenger security will be challenging for SAMoD, with the time of the day playing a essential role since passengers – particularly women \cite{SALONEN2018106} – are more concerned about trips during the night \cite{piao2016public}. In this context, the absence of a human driver is essential \cite{SALONEN2018106, schuss2021letssharearide, fraedrich2016user, Polydoropoulou_2021}. Lavieri and Bhat \cite{LAVIERI2019242} found that not having a driver on board in an AV seems to be particularly problematic for Millennials as they see a driver as a kind of "guardian".
Consequently, confirmed by Biermann et al. \cite{Biermann2020}, there seems to be an increased need for security. As a result, future passengers might tend to accept the use of monitoring systems for the purpose of preventing crime, vandalism, and in case of health emergencies \cite{Biermann2020}. 
In their online survey, Sarriera et al. \cite{sarriera_to-share-or-not} found that major deterrents for adopting SAMoD are being potentially related to unpleasant co-passengers, uncertainty regarding the length of a trip, and preferring privacy during a ride. The authors also discovered biased opinions toward passengers of different social statuses and races, leading passengers to prefer to have more information about their co-passengers \cite{sarriera_to-share-or-not}. An essential factor for adopting SAVs is related to passenger's acceptance to share space and time with strangers \cite{KRUEGER2016343}, which is even more important for leisure trips than business trips \cite{LAVIERI2019242}.

People are still hesitant toward automated systems and giving up their control. \cite{Lundquist1218637} establishes a triad of trust, control, and safety needs to ensure positive user experiences during AV rides. They propose to provide clear communication and transparent system feedback to increase these needs and suggest using visual and auditory feedback about the next stop of automated shuttles \cite{Lundquist1218637}. As mentioned above, the use of monitoring technologies and information sharing has the potential to increase security perception. At the same time, however, it needs to be taken into account that travel information can be a private matter \cite{Brell2019b}.
König et al. \cite{koenig_generationY} evaluated whether information about potential co-passengers influences the acceptability of SAMoD systems and measured how different levels of information affected participants' compensation demands. Detailed information about co-passengers proved to be beneficial \cite{koenig_generationY}. Interestingly, they also found that information about men as fellow travelers resulted in higher refusal rates than information about women travelers \cite{koenig_generationY}. 
In accordance with this observation, women seem to prefer being matched with other women to increase feelings of security \cite{schuss-feministHCI-journal, Polydoropoulou_2021}. Indeed, women-only vehicles already have been discussed in the context of public transportation \cite{abenoza2019women-only-Mexico}, ride-hailing \cite{delatte_user_oriented}, and in the context of SAVs \cite{sanguinetti2019ok}.
The mentioned studies show that potential users have security issues with SAVs and underline the importance for the HCI community to come up with adequate solutions. 

\subsection{Perceived Security, Resilience, and Their Mutual Influence}
Shared driverless travel poses new challenges for resilient system design. Generally, resilience can be referred to as the "process and outcome of successfully adapting to difficult or challenging […] experiences" \cite{APADictionarResilience2023}. In terms of SAMoD, overall system resilience also depends on the psychological resilience of (prospective) users. Particularly passengers' perceived security seems to impact resilience in various ways. Firstly, it influences individuals' trust in the system and their confidence in the system's ability to protect them from potential harm or threats \cite{chiou2023trusting}. This trust acts as a foundation for their psychological resilience, as individuals are more likely to be adaptive and resilient when they feel secure and protected \cite{chiou2023trusting}. In addition, perceived security influences individuals' willingness to report security incidents or vulnerabilities \cite{RHEE2009816}. A culture that encourages open communication and reporting fosters resilience by allowing for timely identification and response to security issues \cite{son2017redefining} – which then again also increases overall system resilience. When individuals perceive that their contributions are valued and acted upon, it enhances their motivation to actively participate in maintaining and improving the system's resilience \cite{doi:10.1146/annurev-environ-051211-123836}. 

Overall, perceived security plays a vital role in shaping resilience by influencing individuals' mindsets, behaviors, and willingness to engage in proactive actions within a system \cite{ahlan2011information}.
In turn, we argue that from a human factors and ergonomics (HF/E) perspective, resilience plays a significant role in influencing perceived security within a (SAMoD) system. When individuals and organizations exhibit resilience, it creates a sense of confidence and trust in the system's ability to withstand and recover from adverse events or security breaches. Taking into consideration the crucial role of perceived security in SAMoD systems (Section \ref{related-wor-acceptance}), this perception of security is vital as it will affect passengers' interaction with and trust towards the system and influence their willingness to use it. Consequently, factors affecting the resilience of both individual users and the overall system should be considered from early design phases. Regarding SAMoD, this particularly involves considering them in the design of suitable user interfaces.

\subsection{Interface Design and Evaluation for Shared Automated Mobility}
SAMoD UIs can range, e.g., from passenger information displays in vehicles and planning and booking applications on mobile devices to terminals at mobility hubs. They will provide the sole basis for the communication of passengers and the intelligent systems as no human operators (e.g., drivers) will be involved anymore. In terms of interaction modalities, already familiar technologies like touchscreens, information displays and control buttons seem to be preferred by potential SAMoD users \cite{Biermann2020}. The preference for established modalities, such as visual and auditory, and interior locations, such as the front area, is reflected by the systematic literature review on the in-vehicle design space by Jansen et al. \cite{designspaces4automatedvehicles}. They provide a comprehensive overview of input and output modalities and information locations and highlight the relevance of multi-modal in-vehicle interactions.

Since most currently available (S)AVs are still limited, prototyping and simulation methods are used to test and evaluate future (S)AMoD UIs. 
To ensure a meaningful transfer of study results to the development of SAMoD systems, it is essential to integrate the highly dynamic element of the context of use \cite{Kray2007, Flohr2021}. In ubiquitous systems like SAMoD, this involves not only the consideration of auditory and visual factors but also the inclusion of surrounding elements like other people that might be present, as well as their relation to the respective users \cite{Kray2007}.  
There are several methods to prototype the physical and social context of human-AV interactions \cite{Flohr2023}, including lab-based prototyping with mock-ups, virtual reality, and simulators \cite{Krome2015, Gerber2019, Flohr2020, Flohr2021, GermanResearchCenterforArtificialIntelligenceDFKI2019}, wizard-of-oz vehicles (WoOz) \cite{Kim2020, Detjen2020, Baltodano2015}, and experimental AVs  \cite{madigan2016acceptance, salonen2019towards, NORDHOFF2018843, Polydoropoulou_2021}. \cite{Flohr2023} provide a detailed overview of suitable methods and discuss the value of context-based interface prototyping for the AV domain.
In general, each method offers advantages and disadvantages that have to be weighed by the experimenters depending on the focus of the study. While the use of experimental AVs for evaluating SAMoD UIs intuitively seems to be the first choice, it needs to be considered that current setups are still quite limited, e.g., to specific test scenarios and low speed limits \cite{NORDHOFF2018843}.If a study purpose can be achieved under these limitations, actual AVs might be suitable. However, to investigate human-AV interactions beyond those limitations, e.g., automated rides in complex urban environments, WoOz and simulators offer promising alternative approaches.
In WoOz setups used to simulate AVs, a hidden driver de facto controls the vehicle, while study participants are told that the vehicle is driving fully automated \cite{Bengler2019}. 

WoOz studies allow for conducting realistic ride studies in complex environments. Still, the method is limited in terms of control and comparisons between rides because each ride varies due to contextual factors (e.g., traffic density, time of day, or the behavior of other road users). In contrast, simulators offer controllable and reproducible test environments \cite{Schoner2015, DeWinter2012}. As \cite{Flohr2020} point out, virtual-reality-based simulators are often quite sophisticated constructs that can be applied to investigate driver-vehicle interactions where the simulation needs to adjust to the participants'/drivers' steering input. Since users of SAMoD systems are passive passengers, it is not necessary to enable study participants to control the simulation. Simulations can thus also be realized using "immersive video" \cite{Kray2007}, which offers a straightforward and time-efficient approach to prototyping SAMoD systems, e.g., \cite{Gerber2019, Flohr2020}. The latter is quite fitting to investigate human-AV interactions in specific situations (e.g., at a particular time of day) in a controllable manner but still with an adequate representation of the dynamic environment.

\section{Material and Method}\label{section:MaterialAndMethod}
To investigate our research question, we created a UI prototype representing a SAMoD in-vehicle passenger information display. Variants of the UI were evaluated in an exploratory within-subjects user study with a diverse sample of participants ($N=24$; gender-balanced, wide range of ages) using a video-based automated vehicle simulator \cite{Flohr2020}. 
With our study, we aim to have a closer look into the information needs of passengers and counteract the limitations of an online survey by simulating rides in an SAV during different times of the day. At the same time, we do not necessarily say this would be the best solution. Quite the contrary, we acknowledge that 1) serious privacy side effects could arise, and 2) stereotypes could be further manifested. Still, we wanted to let participants experience receiving this information on their fellow travelers and discuss with them how it would influence their perceived security. Our motivation was to evaluate whether such a controversial concept would convey security after all and, if so, under what circumstances people would need and want to use it. 

We identified leisure trips as a typical case for the use of SAMoD during both day and night times (more details in section \ref{section:Scenario}). In each simulated ride, an in-vehicle UI prototype (section \ref{section:DesignAndPrototype}) provided participants with information about the ride and fellow passengers. In the respective UIs, we varied the type and amount of information participants received when co-passengers boarded and left the vehicle. Information was either provided \textit{with} personal data on co-passengers (name, age, target destination, profile picture), or \textit{without}.
We used a within-subjects design, and each participant experienced four rides: two night and two day trips, one with and one without personal information about co-passengers. The order in which each participant experienced the variants was randomized and counter-balanced.

Since we wanted to investigate shared rides, we identified two options to include the "sharing" aspect in the study: 1) simulating passengers boarding/leaving the vehicle with sounds and visual information displayed on the UI, and 2) using real persons ('actors') that complement the setup. Regarding the latter, Flohr et al. \cite{Flohr2020} investigated the effect of supplementing SAV simulator studies with actors mimicking co-passengers. While they found some support for the approach, it does not seem to increase participants' immersion in the simulation. Instead, it seems to increase the occurrence of motion sickness symptoms in simulator studies \cite{Flohr2020}. Therefore, considering the potential adverse effect on participants' well-being during the simulator study and the problematic pandemic situation at the time of the study conduct, we decided to simulate co-passengers getting on and off the SAV only virtually. While this supports, on the one hand, our intended focus on the information display, this can, on the other hand, also be considered a limitation of the study, which we further discuss in section \ref{sec:limitations}. 

The study was conducted in accordance with the ethical guidelines stated in the Declaration of Helsinki \cite{2009declarationofhelsinki}. Participants took part voluntarily, were obliged to provide their written informed consent, and had the opportunity to abort the study at any time without stating reasons.

\begin{figure}
  \centering
  \includegraphics[width=\textwidth]{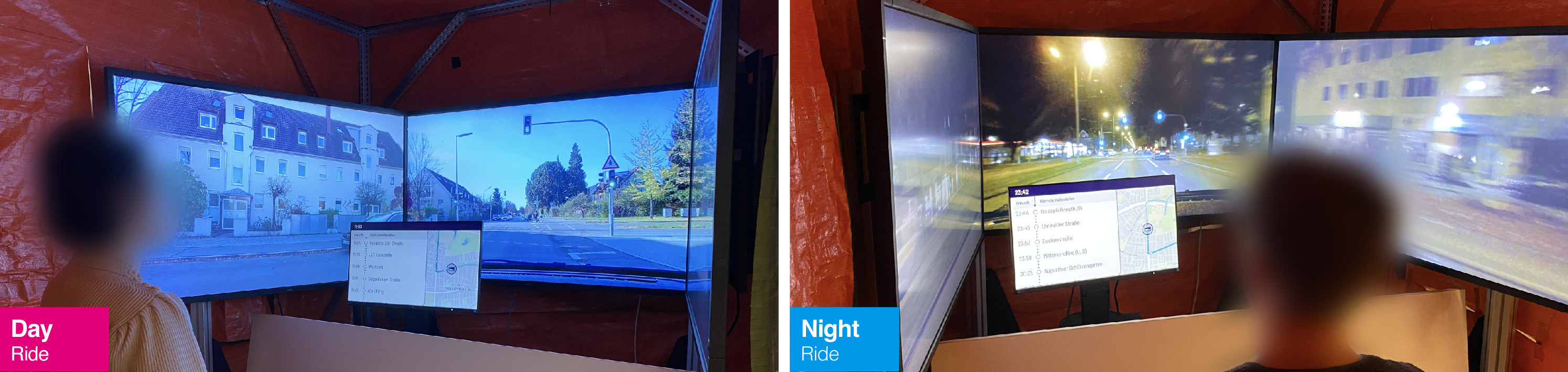}
  \caption{Study participants experienced two day and two night rides in the immersive video-based automated vehicle simulator.}
  \label{fig:teaser}
\end{figure}

\subsection{Setup}
\begin{figure}
  \centering
 \includegraphics[width=\linewidth]{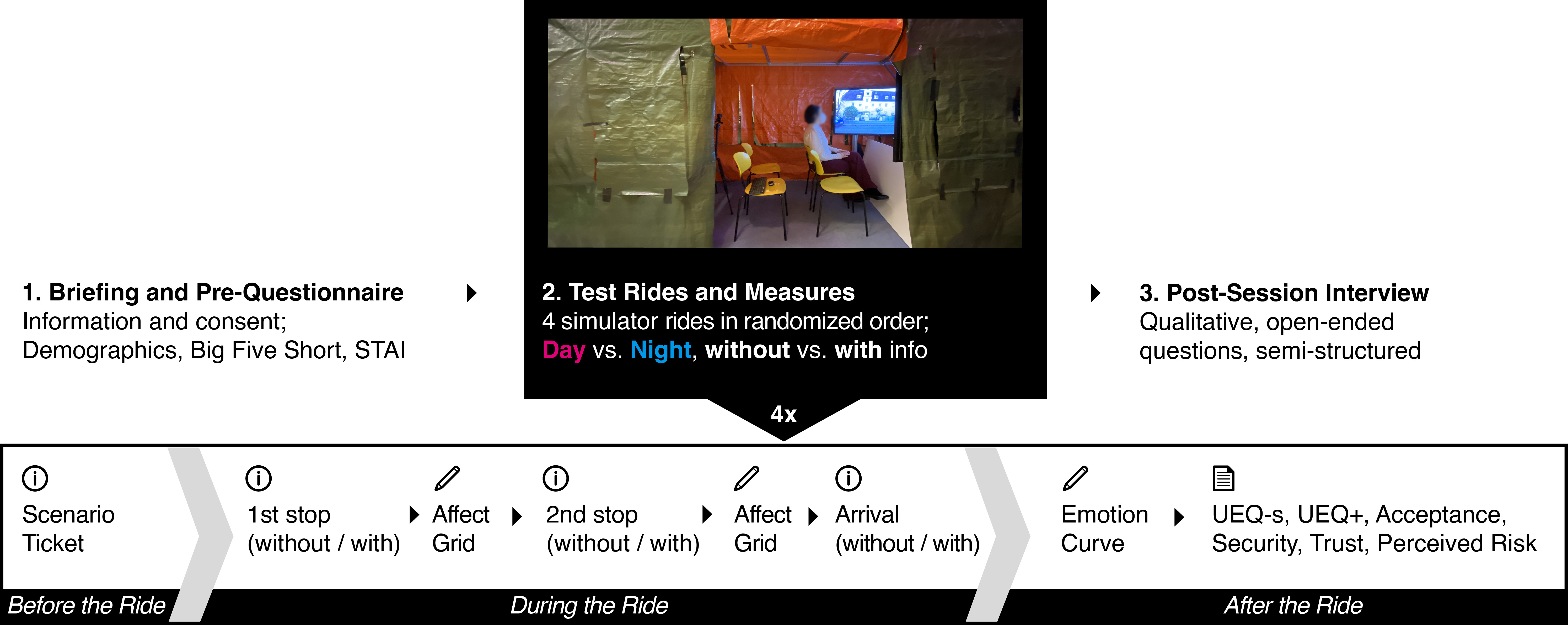}
 \caption{Study procedure (top) and sequence of the four simulated SAMoD rides (bottom).}
  \label{fig:procedure}
\end{figure}

Since contextual factors play a crucial role in passengers' travel experiences and information needs, we intended to establish a realistic but still controllable test environment for the user study. Therefore, we adapted the immersive video-based simulation setup used by Flohr et al. \cite{Flohr2020, Flohr2021} and combined it with a tent-based vehicle mock-up (e.g., used by Schu\ss{} et al. \cite{schuss2021letssharearide}) to provide even more realism. The resulting setup (Fig. \ref{fig:teaser}, \ref{fig:procedure}) consisted of three LCD screens that played back videos representing a passengers' view out of the front, left, and right windows of a shared AV. 

Similar to \cite{Flohr2020}, we used audio and video footage of day and night rides through an urban environment to create simulations for two night and two day rides. The footage was captured using three action cameras mounted in the center of a BMW i3's windshield, as well as on the front side windows. In addition, we enhanced audio footage with additional sounds (e.g., opening and closing noises of sliding doors). Along with a 2x2 seating group, the footage was played back on three NEC Full HD 55.1-inch TV screens situated in a tent-based vehicle mock-up. The tent separated the simulation from the surrounding lab environment to support participants' immersion by entering a closed space when boarding the simulated SAV (Fig. \ref{fig:procedure}). 
The UI prototype of the passenger information display was displayed visually on an additional 24.1-inch screen (Fig. \ref{fig:teaser}). Audio sounds and voice prompts were provided by a Logitech 2.1 sound system.

\subsection{Design Process and Prototypes}\label{section:DesignAndPrototype}
\begin{figure}
  \centering
 \includegraphics[width=\linewidth]{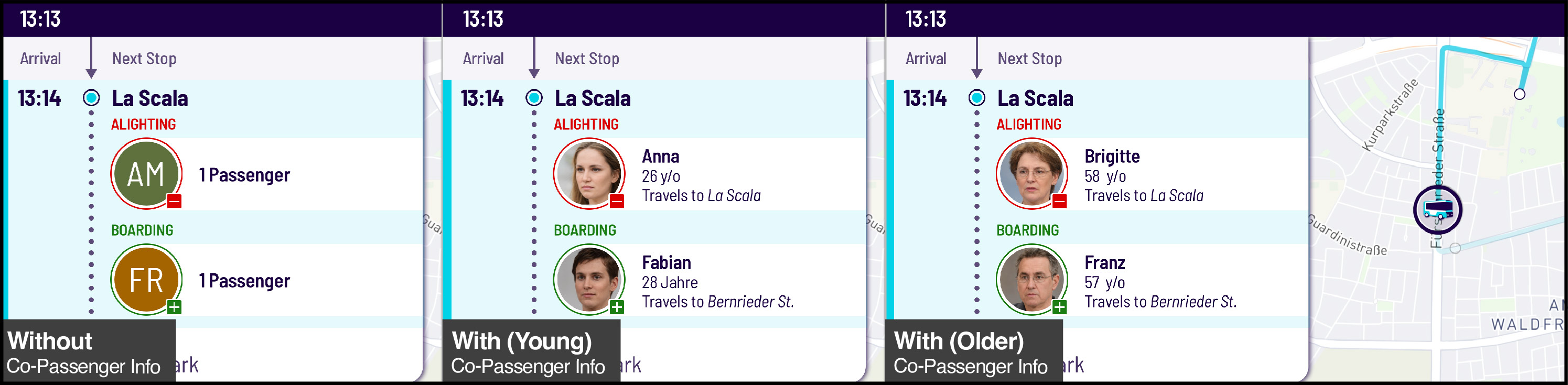}
 \caption{Apart from time of day ("day" and "night"), study conditions varied in the amount of provided information on co-passengers: 1) without information, 2) with information. In the two rides with information, co-passengers' age varied between "young" and "older". 
 }
 \label{fig:UI-design}
\end{figure}

The tested UI prototypes were designed iteratively following findings from related user studies and a comprehensive literature review.
We used video-based prototyping to create high-fidelity visual and auditory UI representations that matched the video-based simulation. The visual information display featured a split-view of 1) a schedule showing upcoming stops, estimated arrival times, and information on co-passengers getting on/off the vehicle, and 2) a map illustrating the current location of the AV and the planned route (Fig. \ref{fig:UI-design}), which follows proposals of previous work (e.g., \cite{sanguinetti2019ok, Flohr2020}).
We created two general prototype variants to investigate the research question (Fig. \ref{fig:UI-design}). While the first variant ("without") does not show personal information about co-passengers, the second variant ("with") features such information by displaying name, age, target destination, and profile picture of co-passengers. For each test ride, participants experienced either a prototype with or without information on co-passengers, i.e., the variant stayed consistent within the rides.

Previous research suggests that combining these data reduces overall compensation demands for sharing a ride with a stranger \cite{koenig_generationY}. We did not include a rating of fellow passengers, as rating systems hold discriminating characteristics \cite{schuss-feministHCI-journal, ruha2019race}. We used AI-generated pictures with neutral facial expressions \cite{generatedfaces} as photos of the entering fellow passengers. We included fellow passengers' age as we hypothesized that this information might influence participants' perceptions. Thereby, we defined two age groups: young (between 20 and 30) and older (between 50 and 60).
Age of fellow passengers was balanced so that each participant experienced one ride with a younger man/woman and an older man/woman as we expected that age could have an effect on passengers' perceived security. 
The provided contextual information (map, street names, etc.) matched the real-world environment where the simulation footage was recorded and animated (using Adobe After Effects CC 2021) according to the simulated vehicle's movements (e.g., the position of the AV in the map). For permutation purposes, we created eight video prototypes of the UI to have one variant without and one with information on co-passengers for all four simulated rides.
Signal sounds and voice prompts complemented the visual UI (e.g., without: \textit{"Next stop: [stop name]. One passenger gets on. One passenger gets off."}; with: \textit{"Next stop: [stop name]. [Name of passenger] gets on. [Name of passenger] gets off."}). Voice prompts were created using text-to-speech conversion by Microsoft Azure.

\subsection{Scenarios}\label{section:Scenario}
We intentionally included participants covering a wide age range in the study, with young people not working yet, and older adults who do not work anymore. To provide for a broad spectrum of participants' real lives, we chose leisure trips as scenarios for the four rides in the study. Since people are reported to be more likely to reject sharing rides with unknown fellow passengers for leisure trips compared to commute trips \cite{LAVIERI2019242}, we wanted to explore whether information about other passengers would mitigate this observation. 
All participants engaged in four trips: two during the day and two at night. We used storytelling to create authentic scenarios for each trip to enhance immersion. The day trips went from a bakery to a park to meet friends and back. The night trips started nearby the passenger's home and had a restaurant as a destination where some friends were supposed to meet and were also round trips.
To get even better acquainted with the scenario, participants received a paper ticket before each ride with their name, destination, departure and arrival time. After reading the scenario to them and handing over the ticket, our participants entered the shuttle bus, chose one of the seats in the front row, and one of the investigators started the video simulation.
During each trip, one man and one woman as a co-passenger entered the vehicle virtually (i.e., this was only stated by the information displayed in the UI prototype). We did not randomize the order, i.e., it was always the woman entering first to avoid losing statistical power due to too many conditions. However, participants always rode with only one person at a time since we hypothesized that it would affect participants' perceived security whether they would be sharing rides with a single man/woman or multiple persons simultaneously. The first (virtual) co-passenger entered at the first stop and got off at the second stop, where the second co-passenger entered the vehicle. At the third stop, participants' reached their target destination.

\subsection{Procedure and Measurements}
We used a mixed-method approach \cite{Creswell2014} and triangulated quantitative data collected during and between rides with observations and qualitative interview data. Each study session can be divided into three parts: briefing and pre-questionnaire, test rides and measures, and post-session interview. Each session took between 60 to 90 minutes in total.

\subsubsection{Briefing and Pre-Questionnaire}
After receiving a briefing comprising general information about the study goal and the procedure, participants signed a declaration of consent. Then, they filled out a demographic pre-questionnaire. 
We also included the short version of the Big Five inventory \cite{bigfiveshort_DE, bigfive_original} to get insights into a participant's personality. Prior research showed that psychological factors and attitudes most likely influence people's adoption of AVs \cite{Yap20161preferences, HABOUCHA201737, zmud2016consumer}. 
As the level of a person's anxiety influences the perceived security \cite{kurani2019user}, we also included the state-trait anxiety inventory (STAI) \cite{STAI-spielberger1983} in our pre-questionnaire. Since current research is not conclusive on whether having experienced any sort of crime has an influence on perceived security \cite{currie2013factors}, we left this aspect out.

\subsubsection{Test Rides and Measures}
During each of the four rides, participants filled out Russell's Affect Grid \cite{russell1989affect} in an adapted emoji-based version inspired by \cite{toet_emoji-affetct-grid} using pen and paper. The Affect Grid is one of the most widespread models for emotion measurement and consists of two dimensions to measure: pleasure (displeasure – pleasure) and arousal (low energy – high energy) \cite{2017_Jeon_HCI_Emotions}.  
Each time information about an upcoming stop and entering or leaving passenger was displayed during the ride, participants were instructed to set a cross to express their current emotional state in the grid.  

After each ride, participants got off the simulated automated vehicle and summarized their subjective emotional constitution throughout the journey by drawing an emotion curve on a template also used by \cite{Kim2020, Flohr2021}. Subsequently, the experimenter accompanied them to a workplace where they filled out a digital questionnaire. Starting with the short version of the User Experience Questionnaire (UEQ-s; 8 bipolar items; 7-point scale; \cite{Schrepp2017}) as well as the Usefulness \cite{Schrepp2019} and Attractiveness \cite{Laugwitz2008, Schrepp2018} dimensions of the UEQ+ (4 bipolar items for each dimension; 7-point scale; \cite{Schrepp2019}) participants assessed their experiences of the ride and respective HMI concept. Since we expected the type and amount of provided information to have an effect on passengers' trust, participants also assessed the Trust in Automation scale of Körber (2 items; five-point Likert-type scale; \cite{Korber2018TiA}). Furthermore, we investigated users' acceptance with the Intention to Use (2 items; 5-point Likert-type scale), and Perceived Usefulness (3 items; five-point Likert-type scale) dimensions of Chen's adaption of the technology acceptance model \cite{Chen2019}. Subsequently, we included Dekker's Security Concerns scale (1 item; 5-point Likert-type scale; 
\cite{dekker2017riding}) and the Perceived Risks scale (1 item; 5-point Likert-type scale; \cite{ribeiro2021customer}) as risk also has an influence on the perceived security \cite{kurani2019user}. 
After the last ride, each participant additionally filled out the Igroup Presence Questionnaire (IPQ, 14 items; 7-point Likert-type scale;
\cite{Schubert2001, Schubert2016}) to assess the quality and immersion of the simulated environment.

\subsubsection{Post-Session Interview}
Finally, we conducted a semi-structured post-session interview with each participant. We asked open-ended questions about the rides in general and the co-passenger information that was provided by the UI. Participants were asked which version of the UI they liked best and why. Participants were also prompted about potential feelings regarding security in the respective conditions, and we inquired whether some information was missing from their point of view. With the consent of participants, audio captures of all post-session interviews were recorded for an in-depth post hoc analysis.

\subsection{Participants}
In total, 24 participants (12 women, 12 men, 0 diverse, 0 n/a; from 18 to 81 years, $M(SD) = 40.5(21.3)$, $Median = 30$) took part in the study. 
All participants were recruited through university mailing lists and word of mouth and attended the study voluntarily. For participation, all of them received financial compensation (approx. 25 US dollars). Their national background was [blinded for review], [blinded for review], [blinded for review], [blinded for review], and [blinded for review]. 
We used the STAI inventory to measure participants' interindividual tendency to evaluate situations as threatening or to react with increased feelings of anxiety. According to the reference values of the trait anxiety scale (items 21-40; \cite{STAI-spielberger1983}) our participants are at the expected medium level of responding with anxiety. The women in our study had a mean value of $M = 36.5$ ($SD = 7.0$; $Mdn = 38.0$; expected value according to references = 37.0) and the men a mean of $M = 35.8$ ($SD = 4.4$; $Mdn = 36.0$; expected value according to references = 34.5).
Participants fall into the average age group between 36 to 65 years and have a high educational level. They correspond approximately to the reference values of the Big5-short (see \cite{rammstedt2013kurze}) for extraversion ($M(SD) = 3.25 (1.29)$; reference: $M(SD) = 3.62 (.91)$), agreeableness ($M(SD) = 3.56 (0.9)$; reference: $M(SD) = 3.43 (.79)$), conscientiousness ($M(SD) = 3.93 (1.03)$, $M(SD) = 3.47 (.95)$; reference: $M(SD) = 4.2 (0.77)$), neuroticism ($M(SD) = 2.45 (0.94)$, $M(SD) = 2.48 (0.9)$) openness to experience ($M(SD) = 3.45 (1.21)$; reference: $M(SD) = 3.70 (0.89)$). We therefore assume that the obtained results are not falsified through a non-representative sample (e.g., a sample with exceptional high scores in neuroticism could have an impact on the perceived security). 

\section{Results} 
For the quantitative results, descriptive and inferential statistics were calculated using JASP 0.16 \cite{JASP2022} and jamovi 2.2.5 \cite{jamovi}.
The audio-recorded post-session interviews were transcribed verbatim and analyzed applying qualitative content analysis \cite{mayring2010qualitative, kuckartz2012qualitative} with MAXQDA \cite{maxqda}. 
Session notes and anecdotal evidence during the study complemented the data collection.

\subsection{Dependent Variables}
In the following, we report on descriptive and inferential statistics for a comparison of the study conditions in terms of our dependent variables, as well as for an assessment of the simulated setup by having a look at participants' presence perception.
We computed repeated measures analysis of variances (RM-ANOVA) to explore differences in the study conditions with the RM factors 'time of day' (day, night) and 'information on fellow passengers' (without, with) as well as the between subjects factor 'gender' (women, men). One woman (P21) only completed three of the four rides due to occurring simulator sickness symptoms. The missing data of P21 was imputed with maximum likelihood estimates (e.g., \cite{Allison2009}) for the respective scales.
When a RM-ANOVA 
returned significant ($\alpha = .05$) for a certain scale, post-hoc tests in the form of Holm-adjusted pairwise comparisons for all conditions were calculated. Effect sizes were interpreted according to Cohen \cite{Cohen1992}.

\begin{figure}[ht]
  \centering
  \includegraphics[width=\linewidth]{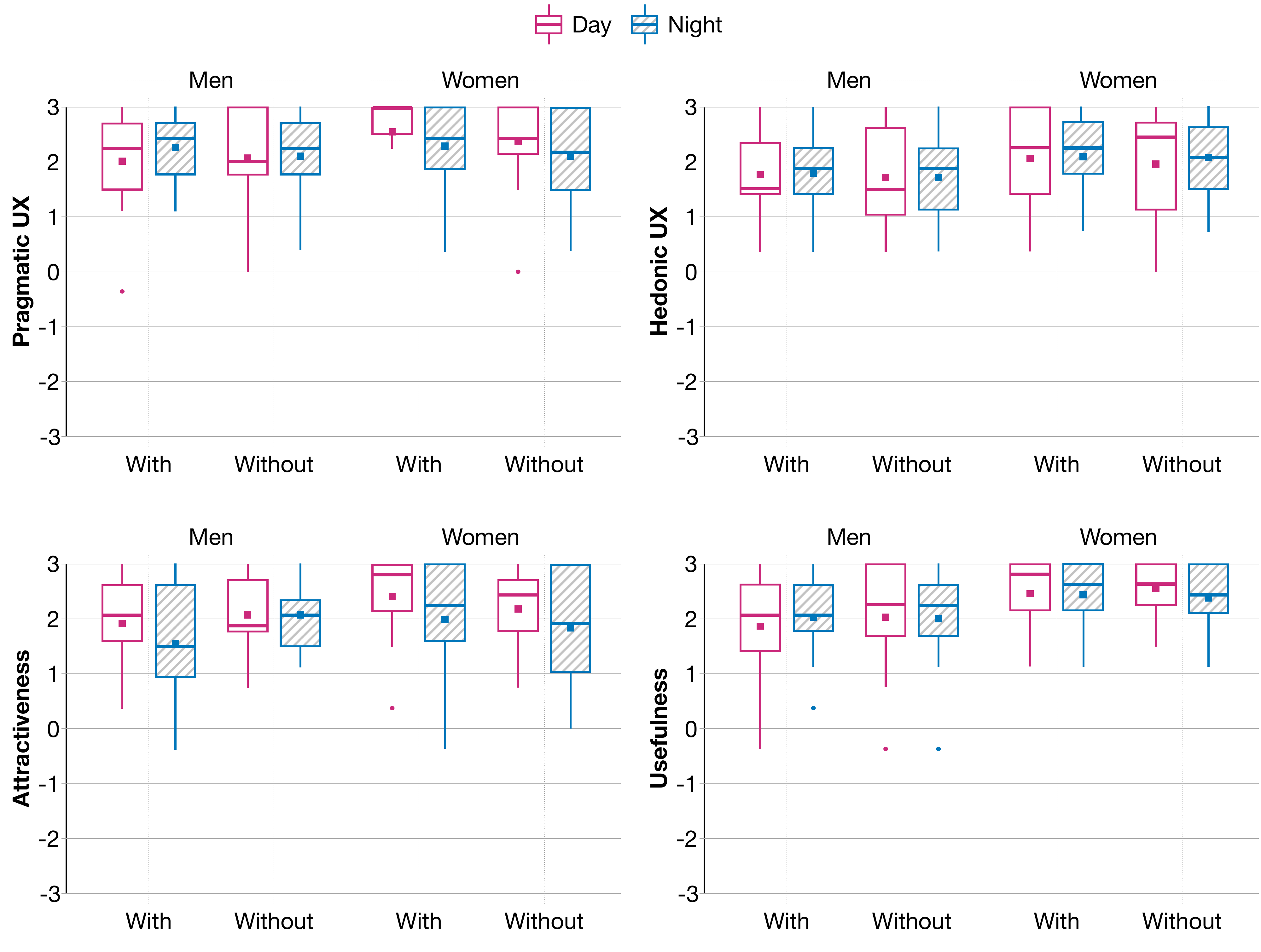}
  \caption{Boxplots of UEQ-s scales (pragmatic UX and hedonic UX), usefulness, and attractiveness (-3 = low; 3 = high) for the four study conditions and the between-subjects factor gender.}
  \label{fig:UX-scales}
\end{figure}

\subsubsection{User Experience}
With reference to the UEQ-s benchmarks \cite{Schrepp2017_UEQBenchmark, Schrepp2017}, the tested SAMoD system received excellent ratings for both pragmatic and hedonic UX quality throughout study conditions (Fig. \ref{fig:UX-scales}). While we did not find meaningful differences in terms of pragmatic quality 
, hedonic quality 
, and usefulness, 
a RM-ANOVA revealed significant differences for the UEQ's attractiveness scale with regard to time of day 
($F(1,22)=6.820, p=.016, \eta\textsuperscript{2}\textsubscript{G}=0.026$) 
and an interaction effect of passenger information and gender
($F(1,22)=5.059, p=.035, \eta\textsuperscript{2}\textsubscript{G}=0.021$).
Post-hoc tests show that participants' overall impression was significantly more positive ($t = 2.612, p\textsubscript{holm}=.016$) during daytime 
than during nighttime (Fig. \ref{fig:UX-scales}), 
with a mean difference of $M(SE)=0.3(0.1)$ and a medium effect of $Cohen's~d=0.533$. Despite the significant results of the RM-ANOVA, an interaction effect of information and gender was not confirmed by subsequent pairwise comparisons.

\subsubsection{Acceptance}
A between-subjects effect of gender returned significant in the RM-ANOVA for both used scales of Chen's TAM \cite{Chen2019}:
Perceived Usefulness ($F(1,22)=7.586, p=.012, \eta\textsuperscript{2}\textsubscript{G}=0.194$) and
Intention to Use ($F(1,22)=6.490, p=.018, \eta\textsuperscript{2}\textsubscript{G}=0.159$). 
Post hoc comparisons confirm that women 
perceive the tested SAMoD system to be more useful than men do (Fig. \ref{fig:acceptance-plots}; $t=2.754, p\textsubscript{holm}=.012$) with a mean difference of $M(SE)=0.5(0.2)$ and a medium-sized effect of $Cohen's~d~=~0.562$. 
Similarly, women 
show a higher Intention to Use the SAMoD system compared to men (Fig. \ref{fig:acceptance-plots}) 
with a mean difference of $M(SE)=0.5(0.2)$ and a medium-sized effect ($t=2.547, p\textsubscript{holm}=.018$, $Cohen's~d=0.520$). 
Apart from the between-subjects effect and the generally medium-high to high acceptance ratings of the SAMoD system, no meaningful within-subjects effects of time of day and passenger information on Perceived Usefulness and Intention to Use were revealed.

\subsubsection{Security, Trust, and Perceived Risk}
\begin{figure}
  \centering
  \includegraphics[width=\linewidth]{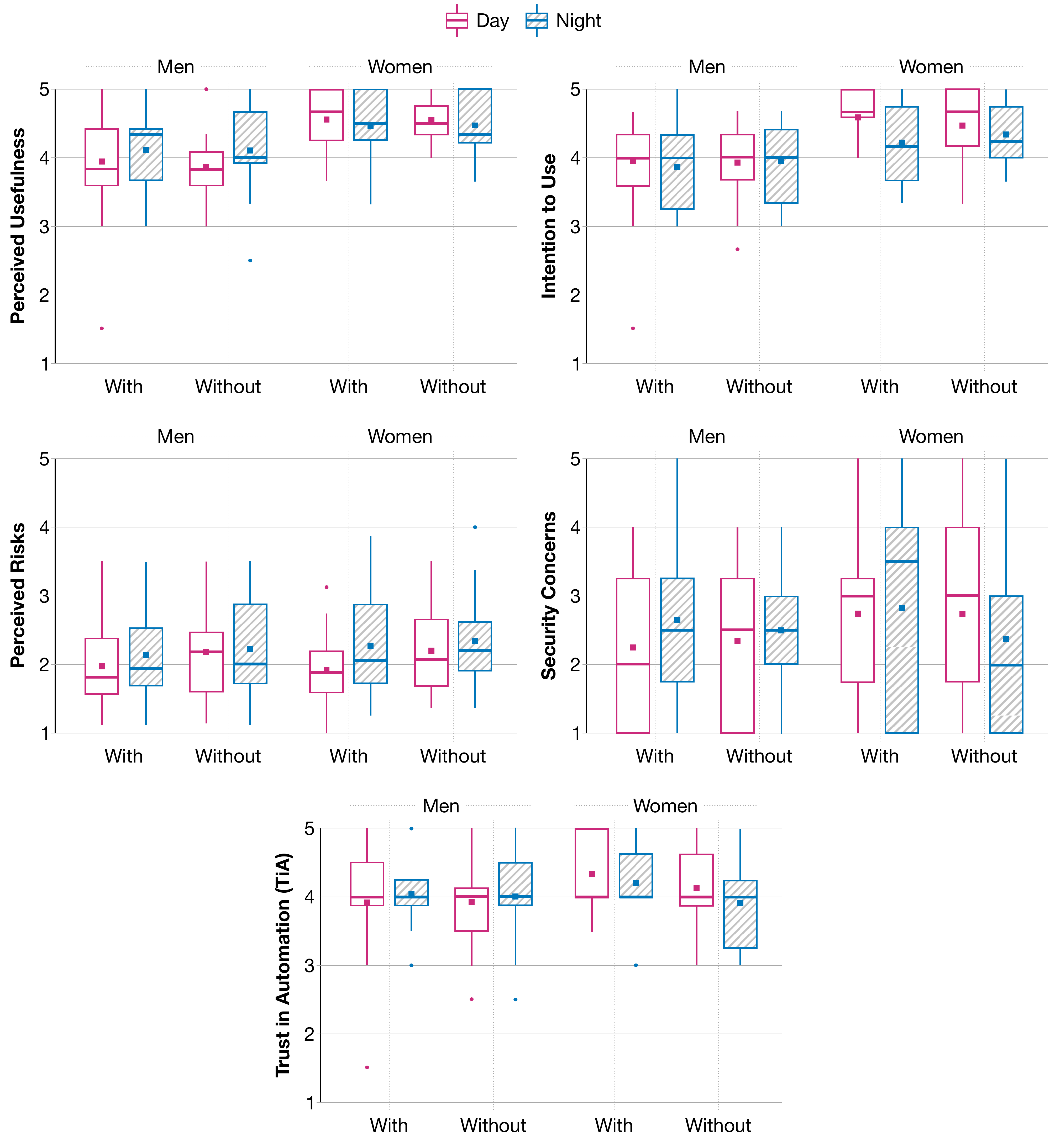}
  \caption{Boxplots of acceptance scales (perceived usefulness, intention to use), trust in automation, security concerns, and perceived risk (1 = low; 5 = high) for the four study conditions and the between-subjects factor gender.}
  \label{fig:acceptance-plots}
\end{figure}

Participants' trust in the automated system was medium-high among all conditions (Fig. \ref{fig:acceptance-plots}).
With regards to the medium-rated security concerns (Fig. \ref{fig:acceptance-plots}),
participants seem to have some, but no severe concerns on their security during their ride. No  meaningful difference induced by time of day or passenger information was detected. 
A significant difference was found in terms of perceived risks ($F(1,22)=7.321, p=.013, \eta\textsuperscript{2}\textsubscript{G}=0.013$). AMoD rides without information 
were perceived as significantly more risky than rides with information about fellow passengers (Fig. \ref{fig:acceptance-plots}) 
with a mean difference of $M(SE)=0.2(0.1)$ and a medium-sized effect ($t=2.706, p\textsubscript{holm}=.013$, $Cohen's~d=0.552$).

\subsubsection{Emotion}
\begin{figure}
  \centering
  \includegraphics[width=0.8\linewidth]{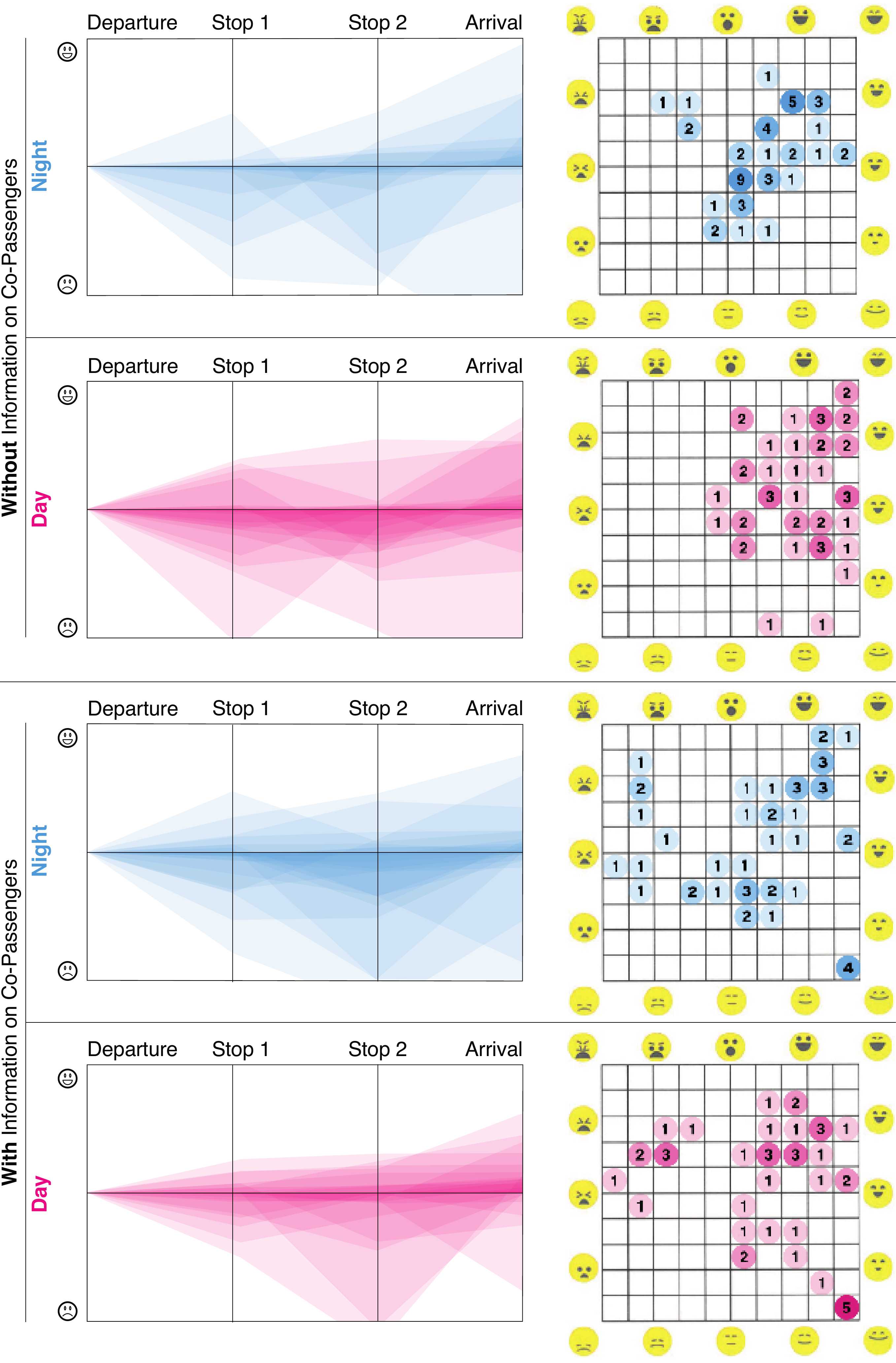}
  \caption{Stacked Emotion Curves (left; Opacity: 0.1; Normalized at ‘Departure’) and Affect Grids (right) for the Four Study Conditions.}
  \label{fig:emotion}
\end{figure}

Judging from visual inspection of the affect grids and emotion curves (Fig. \ref{fig:emotion}), participants found rides during daytime and without information to be most pleasant. Rides without information seem to receive more positive assessments whereas the UI variants with information show higher dispersion in the affect grids. Generally, rides during daytime seem to be perceived more pleasant than night rides.
In accordance with that, the statistical analysis of the quantified ($min=1, max=10$) uni-dimensional subscales of the affect grid (pleasure, arousal) revealed no meaningful effect in terms of arousal but significant differences in the pleasure ratings with regards to the time of day ($F(1,43)=12.386, p=.001, \eta\textsuperscript{2}\textsubscript{G}=0.032$). Rides during daytime ($M(SD)=7.6(2.2)$) received higher pleasure ratings than rides during nighttime ($M(SD)=6.8(2.2)$) with a mean difference of $M(SE)=0.8(0.2)$ and a medium-sized effect ($t=3.519, p\textsubscript{holm}=.001$, $Cohen's~d=0.525$). 

\subsection{Qualitative Content Analysis}
For the qualitative content analysis \cite{mayring2010qualitative, kuckartz2012qualitative}, interview transcripts were initially explored line-by-line. In a second step, we highlighted text passages, searched for keywords, and added notes. Subsequently, the transcripts were scrutinized again and codes were derived from the text by applying inductive coding to refine themes and codes in an iterative process until the final expressions were identified. 
In the following, we present our main findings (e.g., statements expressed during the post-session interviews) with their number of mentions (n) and the number of women and men in our study mentioning them. First, we present the perceptions of the rides in general. Then, we cluster them according to three main topics: information preferences, day vs. night, and the type of information that participants were requesting.
\subsubsection{Presence Perception and Experience of the Rides}

In general, participants described the four rides as positive and considered the ride in the simulator as short, entertaining, and pleasant. Moreover, participants emphasized how realistic the four trips felt to them: \textit{"Yes, it was quite real and I didn’t feel I am in the simulation room and it was so real. It was quite good, yeah."} [P15], which is also reflected in the medium to high ratings for the four IPQ scales (
Realism ($M(SD)~=~4.0(1.1)$), Involvement ($M(SD)~=~3.3(1.1)$), Spatial Presence ($M(SD)~=~4.1(0.9)$), and General ($M(SD)~=~4.9(0.8)$). Participants' immersion in the simulated SAMoD can be judged to be quite high.
Participants compared the simulated AMoD journey to using public transportation systems such as buses or metros today (16; 6 women, 6 men). 

\subsubsection{Information Preferences}
Overall, the qualitative data obtained in the study show that participants favored to have information about their co-passengers (15; 9 women, 6 men) over having no information (8; 2 women, 6 men). The most important reason for preferring the UI version with co-passenger information was security (22; 12 women, 9 men): \textit{"I would have felt more secure with the display with the information and picture."} [P05], \textit{"I felt so much more secure compared to the other version."} [P22].
Participants considered the information as more pleasant (9; 5 women, 4 men) in terms of being connected to others: \textit{"when the person comes in and you have a little info about them, I thought that was pleasant. You could also – in case something happens – address them by name or, yes, it is more pleasant than the anonymous [version]."} [P04]. Other advantages of having knowledge about fellow passengers were that participants considered it to be more interesting (4; 1 women, 2 men) and humane (3; 1 women, 1 men). Participants who preferred having information about other passengers where also willing to share these information about themselves.
In line with this finding, the most important reason for preferring the UI version without information was privacy (17; 4 women, 6 men): \textit{"My first thought was 'Oh no, people will know my name'. I don't like that at all."} [P12]. Other participants regarded this information as not important (4; 1 woman, 3 men).
Displaying passengers' details was even seen as insecure (3; 2 men) or untrustworthy (2; 1 man) as these details could potentially harm people. One participant expressed worries about the security of our young faux passenger ("Anna") as he elaborated: \textit{"Well, at night you're just a bit more insecure, for example, when drunk, young people hop on. So [my worries] were also related to Anna, because people might think 'Oh, here comes Anna now, maybe we can hit on her or something.’ That would be quite insecure for her then."} [P05].
Although of our 24 participants, 15 expressed they would prefer the UI version with information, we would like to point out that this was not a clear decision every time. One participant even was unable to decide which version they preferred. Most participants found pros, as well as cons, for both versions and were weighing these off until finally making a decision. While this reflects how security and privacy are antagonists, the appropriateness of the variants was considered to be highly context-dependent, as outlined in the following paragraph.

\subsubsection{Day vs. Night}
Generally speaking, our qualitative data confirms the difference the time of the day makes for sharing rides in SAVs with strangers as their number of mentions is higher (35; 9 women, 7 men) than statements that do not emphasize this importance (9; 2 women, 7 men). 
In this context, participants stated time-related concerns like \textit{"during the night one is generally more careful and feels vulnerable} [P09]. Several women (17; 9 women) expressed concerns when sharing rides with unknown men and said they would favor sharing a vehicle with other women at night over mixed vehicles. For instance, [P03] explains \textit{"well, especially in the dark. During the day is not that tragic, but in the dark, I don't want to share a ride with a man or get off the vehicle with him."}.
Interestingly, some of the men in our study conveyed similar feelings towards sharing rides with other men (8; 7 men) – particularly at night: \textit{"Because it was Brigitte who got on at the first stop and then at the second stop it was a gentleman. That indeed made a difference to me."} [P11]. As a reason they stated to feel more secure as a statement like, i.e., \textit{"men tend to be more aggressive"}[P05] indicates.
Participants also made clear that they would not need the displayed information on co-passenger during the day, but would prefer to have the information during the night: \textit{"Especially at night it was more pleasant for me and more important. [...]. The fact that I was registered, for example, the [man/woman], as well. Yes, that was much more important for me at night than during the day."} [P22].
\subsubsection{Type of Information}
We asked participants which type of information they considered to be the most important one/s. The fellow passenger's profile picture was regarded to make all the difference (23; 7 women, 8 men) since it gives \textit{"an impression of the person that is going to get on the vehicle at a glance"} [P16]. In this regard the photo seem to give participants a feeling of control over the situation while the other information provided was rather a \textit{"nice-to-have"} [P23]. Knowing beforehand who would enter the vehicle also conveys security: \textit{"Yes, I mean, I saw the picture and it looks nice and I actually had less fear."} [P03].
The co-passenger's gender was essential, as well (13; 5 women, 2 men), followed by their age (10; 6 women, 2 men), the name (8; 3 women, 2 men), and the respective destination of the co-passengers (7; 2 women, 2 men). 
Most of the participants in our study stated that the information the system was offering was sufficient and emphasized how helpful it was to see the vehicle's route and its arrival time on the display. Some participants provided improvement suggestions such as getting information in case people with special needs, big luggage, or strollers would enter the vehicle, or whether seat belt use was compulsory.
\section{Discussion}
Overall, the results underline people's openness towards SAMoD, which is in line with previous work \cite{schuss2021letssharearide, riener2020autonome}. Participants considered SAMoD to be useful and reported relatively high trust in the technology, intention to use, and positive experiences of the (simulated) SAMoD rides. However,  participants also expressed concerns regarding security – especially with regard to night rides. In the following, we discuss our findings in detail and situate them among previous work.

\subsection{Night Trips Require Higher Levels of Information}
In general, the SAMoD rides during the day were evaluated more positively than night rides. Participants consider the overall attractiveness of the SAMoD system higher and report more pleasant rides during the day. Rides without information about co-passengers were perceived as more pleasant than rides with information. We hypothesize two reasons as sources of this findings: 1) participants are used to receiving no information about others when sharing a ride (as is the case in public transportation), and 2) people generally prefer rides during the daytime. This interpretation is comprehensively supported by our qualitative data and is in line with existing data from research in public transportation \cite{piao2016public,itf2018womenssafetypublictransport, 2010_Blom_FearandtheCity}. 

In contrast, rides with information provided by the in-vehicle UI were experienced to be significantly less risky compared to rides without information about co-passengers. 
Again, this is reflected in our qualitative data, with 21 participants underlining increased perceived security through the information. This can be taken as a general preference for information about co-passengers — particularly during the night and is in line with \cite{IMWUT_bystanders}, who found that people are willing to provide information such as their gender, age, etc. to visually impaired persons in public spaces, if higher security assurances can be made.
While during the day, information about fellow passengers seems to have rather adverse effects (e.g., in terms of emotion), this changes during the night, where it has, on the contrary, positive effects. 
Prior work underlines the importance of privacy particularly in public transportation \cite{IMWUT_privacy}.
Security and privacy are often antagonists in today's public systems and this dynamic has implications for resilience from a human factors perspective. This became evident in our study as participants mentioned privacy concerns when displaying personal information about other passengers, or themselves. During the interviews, participants weighed the pros and cons of having (no) information. Despite a preference for information during the night, this was not a clear outcome, which is also apparent in a higher dispersion of the Affect Grid assessments for the rides with information.
 While the information on co-passengers positively influenced security for some participants, there were also concerns that this information could have a negative effect exactly on security as strangers would know one's name and destination. To overcome the conflict between security and privacy, it needs to be investigated which information people feel comfortable sharing in order to increase perceived security.
 
 From a human factors perspective, it is crucial to design systems that allow for individual differences and preferences, considering passengers' diverse needs and concerns. Resilience can be fostered by providing customizable options for displaying personal information, allowing passengers to make informed choices that align with their comfort levels.
\subsection{Both Men and Women Prefer Sharing Rides with Women}
While both men and women generally considered SAMoD systems useful, women rated them significantly more so and uttered a higher intention to use such services. We assume that finding is related to their (security) concerns in today's public transportation systems, especially considering night rides \cite{schuss2021letssharearide, schuss-feministHCI-journal, piao2016public, SALONEN2018106}. In combination with the qualitative data and the discovered interaction effect of passenger information and gender, this finding provides evidence that women seem to consider SAMoD systems as more secure than 'classic' public transportation. 
Women and men alike explained in the interviews that they prefer sharing rides with women. This is in line with the findings of \cite{koenig_generationY}, who found people have higher refusal rates towards men as co-passengers. On the other hand,  Polydoropoulou et. al \cite{Polydoropoulou_2021} found different preferences of passengers for sharing with women/men between countries and cultures and that the number of fellow travelers further influences those preferences. In our study, we focused on rides with only one co-passenger as we expected this constellation would have the biggest effect on security. However, our results and the results from previous work \cite{koenig_generationY, Polydoropoulou_2021} underline once more the complexity of the topic.

\subsection{Balancing Security and Privacy as a Design Challenge}
In terms of overall SAMoD system design, considering resilience from a human factors perspective is crucial. However, there is most likely no 'one-fits-all' solution \cite{mirnig2020capacity}. As, e.g., passengers' security needs are higher during the night, our data points toward flexible solutions for different times of the day.
Based on our results, we propose that UIs for ride-sharing should provide general information on the route, arrival time, subsequent stops, and further information and functionalities to increase passengers' (feeling of) security for night rides. 
Providing information on fellow travelers can serve as a suitable option to do so. In our study, having a photo of fellow travelers was considered the most important information unit and was beneficial for passengers' feeling of security, while information on age, name, and destination played a subordinate role. 
In the study, we chose portraits with neutral facial expressions. However, other expressions might induce different – positive or adverse – feelings, e.g., feelings of insecurity. 
Given that photos seem to provide passengers with (at least some feeling of) control over the situation, they might be used in booking apps or in-vehicle displays. 
Passengers could then look for an alternative vehicle, or leave the vehicle at the next stop if someone's photo would make them feel uncomfortable. The feeling of control has been shown to have a positive effect on psychological security in the context of public transport \cite{Gerhold_Sicherheitsempfinden} and, based on our results, we hypothesize that displaying a photo fosters this control, aligning with the principles of resilience.. 
However, given the disagreement among our participants and the aforementioned privacy issues, we suggest 1) not exposing sensible data about co-passengers during the ride and 2) considering alternative approaches. In terms of (1), it might be beneficial to relocate the information retrieval about fellow passengers to another time and place, e.g., the booking phase. For instance, \cite{koenig_generationY} compared private and shared options on a mobile booking app and found that people tend to rather opt for shared rides when having detailed information on their fellow travelers prior to booking. This could also serve as a means to increase (perceived) security. In terms of (2), Schuß et al. \cite{schuss2022youllneverridealone} propose a "buddy system" to address women's security needs (during the night) that takes advantage of the fact that other passengers can also provide security. Instead of seeing them as potentially harmful, their approach focuses instead on the fact of not being alone and feeling secure instead of the feeling of controlling the situation through information. The concept of "social passengering" \cite{imwut_socialpassengering} among passengers inside the same or different vehicles points to a similar direction and might be beneficial for the perceived security.
 By acknowledging the trade-off between security and privacy and offering flexibility in information disclosure, SAMoD systems can adapt to individual needs, enhancing passengers' overall experience and resilience within the system.

\subsection{Limitations}\label{sec:limitations}
SAMoD is still a relatively 'theoretical' subject \cite{Philipsen2019} with real-life applications remaining missing. Therefore, we let our participants experience a SAMoD system in a simulated environment. While participants report high presence perception and immersion, external validity is impaired due to the lab-based setup.
As we were weighing off the negative side effects that come with lab studies, we opted for the simulated environment over conducting, e.g., a WoOz study in real traffic conditions, to compare the study conditions while ensuring high internal validity and high controllability. 

As mentioned in section \ref{section:MaterialAndMethod}, we decided to simulate the presence of other passengers in a shared ride only virtually with sounds and display visualizations. While this was in line with the recommendation of \cite{Flohr2020} and facilitated the study's conformity with applicable hygiene regulations during the Covid-19 pandemic, the representation of a shared ride's social contextual  is limited. On the other, considering our study design with multiple measurements during a test ride, the physical presence of another person might have affected participants' assessment of the information and consequently the study's reliability. Furthermore, we did not intent to focus on the inherent social factors or mutual relationships (that definitely  play an essential role in the context of shared mobility), but focused on the provided information. 
Nevertheless, this should be considered when conducting further studies on SAMoD.

Taking into account the large and diverse population of future SAMoD users, our study has been conducted with a small sample and, although having placed value on a broad spectrum of people (gender-balanced, different age groups, different cultural backgrounds, different education levels), it covers only a part of the variety of potential users. According to the STAI inventory, our participants had relatively low levels of trait anxiety. Since this trait likely has an effect of risk and security evaluation, generalizability is limited. 

Furthermore, the study was conducted during the COVID-19 pandemic. We applied precautions like distancing and hygiene measures and followed the regulations of local and national authorities. While we consider the pandemic's effect on the study conducted to be minor, it might have affected the sample composition as, e.g., only people with medium fear and anxiety have signed up for the study.
It would be interesting to repeat this study with people that show higher levels of trait anxiety as this trait influences the evaluation of risk and security of situations, and we hypothesize that these people could have evaluated the presented prototype in a more positive way. 

The selection of the displayed information on co-passengers covers only a part of the potential variety and might have fostered stereotypes. 
We derived the solution with information about co-passengers based on existing research findings \cite{schuss2021letssharearide, koenig_generationY} and aimed to evaluate whether the availability positively influences security, UX, trust, and acceptance of SAMoD passengers.
By no means we intended to manifest potential stereotypes or the exclusion of people through our selection. However, we want to point out that the selection likely affects the results (e.g., people might refuse rides with others due to their "look"). We are aware that the gender and age of other passengers is a limited view. Other factors, such as race, appearance, or the supposedly associated social statuses definitely play a role in people's assumptions about other people. However, we did not include more personal characteristics to 1) not confound too many different independent variables in the display variants and 2) we wanted to draw a clear line between evaluating the information about other passengers and participants' potential biases about, e.g., other cultures, as we aimed for the former.

\subsection{Future Work}
Passengers' information demands in SAMoD systems are a highly complex and context-dependent issue requiring more research, especially on how to overcome the conflict between security and privacy by design.

Based on our results, we suggest extending the conduct of context-based empirical studies investigating factors like daytime and fellow passengers in SAMoD systems along the whole travel journey. Since, e.g., security issues are relevant for the booking, the ride itself, and on-/off-boarding \cite{schuss2021letssharearide}. While our study focused on the ride itself, further (empirical) studies should also consider the booking phase and the off-boarding when investigating the effect of co-travelers and time of day on passengers' need for information and controls. Here, additional information and safety measures (e.g., emergency/support button) might support passengers' feeling of control and security.
To yield results with high external validity, future studies might include more contextual factors such as the (physical) presence of various and multiple other people in SAMoD rides during different situations. E.g., actors could be used to mimic specific situations \cite{Flohr2020}. 

It would also be interesting to repeat this study in different cultural contexts, as we conducted our study in Germany, where security in public transportation offers high levels of security \cite{Gerhold_Sicherheitsempfinden}. However, we assume that conducting similar studies in countries, such as India or Latin American countries, where public transportation is more difficult to access  – especially for women  \cite{itf2018womenssafetypublictransport} – might yield different results. The applied simulation environment presents a context-based prototyping approach that can be used, e.g., to replicate this or similar studies in other countries and investigate potential cultural differences regarding passengers' (information) requirements. 
Future work might also consider the potential impact of culture and race as an independent variable in the information display. This could result in an exploration of people's explicit and implicit biases based on given prior knowledge.

We used the front of the vehicle as the output location of the information, as these are common modalities \cite{designspaces4automatedvehicles}. Future concepts might also investigate whether (the combination with) other modalities, such as tactile, influence the perception of the presented information and the feeling of security.

\section{Conclusion}
In this paper, we report on a simulator user study ($N=24$) investigating the effects of time of day and provided information on fellow travelers on SAMoD passengers' UX, acceptance, feeling of security, and emotions in shared automated rides. While the evaluated SAMoD system received excellent assessments of hedonic and pragmatic UX, trust, and acceptance, participants emphasized security concerns – mainly when using SAMoD at night. Furthermore, both women and men preferred sharing rides with women over sharing rides with men as co-passengers during the night, whereas, during the day, this information negatively affected participants' evaluation of the SAMoD system.
Associated risks were experienced lower when participants were provided with information about their co-passengers. Most participants generally preferred having information on co-passengers, with photos of fellow travelers considered the most important information element.
However, our results yield ambiguities since providing personal information also triggered privacy concerns among participants. This can be taken as an illustration of the complexity of psychological security and its context dependency. 
Building upon these findings, providing UIs with information on fellow passengers can support SAMoD passengers' feeling of security in shared rides and potentially improve UX, user acceptance, and overall system resilience. However, due to privacy concerns and associated risks, the timing and placement of the information need to be questioned. It might be beneficial to provide this information during the booking phase but not within the vehicle. Future work should consider the whole travel journey of SAMoD, foster the inclusion of contextual factors, and investigate how the provision of additional information and safety measures (e.g., emergency and support features) can increase passengers' feeling of control and security.

\section{Acknowledgements}
This research was partly funded by the German Federal Ministry for Economic Affairs and Climate Action (BMWK) under grant number 19A21047I (SUE) and by the German Federal Ministry of Transport and Digital Infrastructure (BMVI) under grant number 16AVF2134G (APEROL). 
We want to thank Claus Pfeilschifter for the technical support and Tatjana Röhr for the support in the study conduct and data analysis.
Furthermore, we want to thank the study participants for their participation and the anonymous reviewers for their time and helpful feedback.




\bibliographystyle{elsarticle-num}
\bibliography{main.bib}







\end{document}